\documentclass[12pt]{article}

\begin{document}
\def\thebibliography#1{\section*{REFERENCES\markboth
 {REFERENCES}{REFERENCES}}\list
 {[\arabic{enumi}]}{\settowidth\labelwidth{[#1]}\leftmargin\labelwidth
 \advance\leftmargin\labelsep
 \usecounter{enumi}}
 \def\newblock{\hskip .11em plus .33em minus -.07em}
 \sloppy
 \sfcode`\.=1000\relax}

\hoffset = -1truecm
\voffset = -2truecm


\title{\large\bf
Mathematics : The Language of Science}
\author{
{\normalsize\bf A.N.Mitra \thanks{e.mail: ganmitra@nde.vsnl.net.in
}Dept of Physics, Univ of Delhi,  Delhi-110007,  INDIA
 }}

\maketitle

\begin{abstract}
 The purpose of this
essay  is to bring out the unique role of Mathematics in providing a
base to  the diverse sciences  which conform  to its  rigid
structure.  Of these the physical  and economic sciences  are so
intimately linked with mathematics, that they have become almost a
part of  its structure under the generic title of Applied
Mathematics.  But with the progress of time,  more and more branches
of Science  are getting quantified and  coming under its ambit.  And
once a branch of science gets articulated into a mathematical
structure, the process goes beyond mere classification and
arrangement, and becomes eligible  as a candidate  for enjoying  its
predictive powers ! Indeed it is  this single property of
Mathematics which gives it the capacity to predict  the  nature of
evolution in time of the said branch of science.  This has been well
verified in the domain of physical sciences, but now even biological
sciences are slowly feeling its strength,  and the list is
expanding.
\end{abstract}




\section {Introduction}

 : " I think, therefore I am ". -- Rene de Cartes

Mathematics has ben so much ingrained in  the  very thinking of
Mankind since the days of  Plato, Aristotle and Ptolemy, that it is
hard to offer a formal definition for this  unique creation of
Nature. Nevertheless some great thinkers have attempted  approximate
descriptions to capture its essence.  Thus, according to  Bertrand
Russell,  "Mathematics is  the chief source of the belief in eternal
and exact truth, as well as a sensible intelligible world ". But
such an  omnopotent view of  Mathematics is not shared by all
thinkers. For  another giant (Goethe)  felt otherwise:  "Mathematics has
the completely false reputation of yielding infallible conclusions".
Eugine Wigner was dumbfounded by  "the unreasonable effectiveness
of mathematics", yet felt an "eternal gratitude"  for the same.
Despite  such  disparate views  on  the unusual
powers of Mathematics,  there is almost universal agreement on the
unique role of Mathematics in shaping  mankind's thinking on diverse
phenomena of nature. Therefore  the universal appeal of Mathematics
as  the  language  of  Science  -- the subject of this essay -- will
probably strike a concordant note with anyone interested in  exploring
its dimensions.
\par
To give a broad analogy, the position of Mathematics vis-à-vis the
Sciences has been likened to that of the main trunk constituting the vast Tree of
Knowledge, while the Sciences occupy positions corresponding to the
different branches sprouting successively outwards in decreasing
order of theoretical basis. Thus the physical sciences --more
especially physics--correspond to the main branches adjoining the
trunk, while the various applied sciences and their derivatives,
biological sciences, social sciences (especially economics), and so
on, branch further and further out on this Tree. This looks like a
working model for putting in perspective the role of Mathematics,
not only in its own right, but also in shaping the various branches
of Science which -- once put under its ambit--must automatically share its
logical basis. And once you have succeeded in putting  your physical
premises within a mathematical framework, you may rest assured that its
huge dynamical powers are freely available to you for predicting
the outcome of your investigations in more directions than one,
something  your physical  intuition alone  was  utterly
incapable of  anticipating. On the other hand, not all aspects of
the model--and even a formidable one like Mathematics is no
exception--can be taken literally, lest the oversimplified
conclusion of a model being a substitute for reality,
should  obscure our thoughts.

Some aspects of this simple model are convincing enough. For
example, it is a fair statement that, just as the trunk is a more
rigid structure than the branches, so is the fabric of mathematical
reasoning stronger (and tougher) than the flexible format of
reasoning in physics. Indeed if Mathematics is structured on the
strong and short-range forces of purely deductive logic, physics may
be thought to be held together by the (weaker) long-range forces of
analogy, intuition as well as observable evidence. But the quest for
a `mathematical proof' of a successful physical theory which is
concerned with `deciphering the secrets of nature'--often by
unorthodox means--is not a properly defined exercise. After all, such
" proofs "  cannot be more convincing than the inputs on which the
physical theory is based in the first place, and the latter derive
their support from various indirect evidences which have no place in
a formal mathematical theorem. In `pure'  mathematics on the other
hand, there is no place  for any hypothesis / hypotheses other than
those that are present in the statement of a particular theorem.
This is just as true of a simple Euclid's theorem as of the more
complex Yang-Baxter theorem [1]. On these premises it is not
difficult to imagine that a fool-proof mathematical theorem is not
necessarily a good physical theory, especially if  its `hypotheses'
do not have adequate physical support, or vice versa. A famous
example of this apparent paradox is Heisenberg's theory of
turbulence which was "proved " by a mathematical theorem to be
"wrong" , and yet was found to be in excellent agreement with
experiment. This story was told by Werner Heisenberg in a lecture
arranged by Abdus Salam at the International Centre for Theoretical
Physics (Trieste) in 1968 [2], which was presided over by Paul Dirac.
And this work represented the content of Heisenberg's Ph D thesis
carried out under the direction of his teacher Arnold Sommerfeld who
had insisted that his student should  rather do some  `solid' work
for a Ph D  than indulge in some `airy' ideas like matrix mechanics
which was apparently too "`speculative"' to risk for a doctorate!

\section{Pure vs Applied Mathematics}

Nevertheless most physical sciences have fairly well-defined domains
of jurisdiction  characterized by definite procedures for
formulation of problems, as well as elaborate techniques for
solution, a scenario in which Mathematics is both an indispensable
tool  for procedure as well as an essential language of description.
Indeed, many of the physical sciences, especially mechanics,
elasticity, fluid dynamics, magnetohydrodynamics --and even the
General Theory of Relativity for that matter--have grown out of a
deep involvement of mathematicians in these fields  which by usage
and tradition were once regarded as  different domains of Applied
Mathematics. In contrast, the more traditional branches of physics--
theoretical physics, astrophysics, and quantum mechanics--have
generally been regarded as belonging to the physical sciences,
despite deep involvement of mathematicians in these fields.  These
anomalies reveal  an artificial kind of barrier between the
domains of  mathematics and physics,  which  has more to do with the
history of usage than any serious logical reasoning. In particular,
as the physical sciences have evolved together with their
associated experimental programmes , those topics which once were
thought to belong to Applied Mathematics have inevitably shifted to
well-defined areas of  physics and physical sciences.  Perhaps  the
only two subjects which are still  thought to belong to Mathematics
proper--albeit in applied form--are Statistics as well as its
thermodynamic counterpart in  Statistical mechanics. They generated
their own momentum, thanks to  the seminal contributions of
Boltzmann and Planck and Einstein and Smoluchowski, and have stayed
active ingredients of mathematics.  Apart from "`owning"' these
subjects,  Applied Mathematics has largely stayed content with
providing a "temporary  shelter" for many branches of Science which
were once  found to be amenable to the logic of Mathematics , but
eventually developed into well-defined disciplines on their own,
 albeit with a strong mathematical orientation. This is
particularly true of subjects like mechanics, elasticity and  fluid
dynamics to name some, which have long remained under the ambit of
Applied Mathematics-- perhaps for lack of enough observational
motivations --, but there are now distinct signs of at least some
of them (especially General Relativity)  branching out into
independent disciplines on the strength of observational motivations.

\subsection{Hardy's "`apology"'}

Concerning the place  of   Applied Mathematics vis-à-vis Pure
Mathematics,  within the framework of Mathematics as a whole, the
famous mathematician  G H Hardy, in his book `A  mathematician's
Apology' [3] has helped greatly  in putting this issue in a clearer
perspective. Hardy makes the following points (in his own words) :\\
1)  I said that a mathematician was a maker of patterns of ideas,
and that beauty and seriousness were the only criteria by which his
patterns should be judged. \\
2)  It is not possible to justify the life of any genuine
mathematician on the ground of the utility of his work. \\
3)  One rather curious conclusion emerges, that pure mathematics is
on the whole distinctly more useful than applied. \\
Hardy's perception of pure mathematics is quite unambiguous insofar
as the definition of the `core domain'  is fairly absolute, and does
not depend on possible interactions with other fields of knowledge.
His justification of the life of a genuine mathematician is
strangely reminiscent of Michael Faraday's  remark on the " use of a
new-born baby" , in response to a query on the possible utility of
his discovery of the law of electromagnetic induction. His
definition is also not inconsistent with the `tree-trunk' analogy
which merely emphasizes the feeder role of Mathematics for the
development of the other sciences. Perhaps the biggest asset of
(pure) Mathematics is its capacity to predict an  outcome by virtue
of its closely knit logic. For,  once a branch of science gets
articulated into a mathematical structure, the process goes beyond
mere classification and arrangement, all the way to the fruits of
its  predictive powers ! Indeed it is this single property of
Mathematics which gives it the capacity to predict  the  nature
of evolution in time of the said branch of science. And since this
power stems from its logical structure, the issue centres around
the very process  of mathematical thinking which in turn
presupposes  the existence of \textit{order} as the very basis of
mathematical logic.  In this respect it has been  argued [4]
that the boundary between order and disorder is
the realm of reason, the playfield of  creativity ! As to the
precise relation of Mathematics  with  creativity, however, a formal
consensus seems to be lacking.

\subsection{Relation to the sciences: role of `identity' }

To come back to the role of Mathematics in shaping the sciences, Hardy's
reluctance to give a precise  status to Applied Mathematics perhaps
stems from the absence of an `identity' (a core definition)  akin to that
of Pure Mathematics which he had himself  provided. This is probably
because of the mere `umbrella' role of Applied Mathematics in providing
temporary shelter to newly emerging disciplines  with a strong mathematical
flavour but whose pace of development was not  significant enough
to let them claim an identity (independent status)of their own.  And because
of its mere  umbrella status, Applied Mathematics has missed  a
formal identity which most other sciences (physical, biological,
economic) enjoy by virtue of their independent sources of inspiration.
The former has stayed content with merely providing a
`jacket'  for the developing sciences needing its language
and tools. In the process new mathematics has often got created,
even though the science concerned has enjoyed an identity independent
of  the Mother Science (Mathematics !) whose job is only to create fresh
mathematics  for its own sake. This has frequently happened in the
domain of Theoretical Physics  where historically, `new` mathematical
patterns have  often got created, although the source of inspiration
had stemmed entirely from  within. In other words, while  the task
of the mathematician is to create new mathematics \textit{per se},
the task of the physicist is to determine in which domains of physics
these new creations of mathematics should apply. But there is nothing
in the books to  prevent any interaction between the two disciplines,
since each provides inspiration and motivation to the other, often
blurring their mutual dividing line.  Indeed this interaction has
sometimes been so strong as to give rise to the term
"mathematical physics" to represent this mutuality. And the race for
new mathematics for its own sake has been particularly noticeable
in the further development of theoretical physics --
quantum field theory and  String theory-- where the original
physical motivation for understanding new phenomena got completely
lost in the enthusiasm for mathematical self-consistency per se.

\section {Mathematics vs Physics : A special relationship}

In view of the close historical link of Mathematics with
Physics,  it is tempting  to dwell further on the extent of this
relationship through some leisurely examples, the very first one
being  Newton's Laws of Motion. Indeed it was
to give mathematical shape to these laws--especially the Second--that
Newton had to invoke a most vital branch of  Mathematics, viz.,
Differential Calculus,  first discovered by the German mathematician
Liebnitz. This language proved so elegant and so versatile that it
became amenable to elaborate formulations at the hands of great
thinkers like Laplace, Lagrange, Gauss, Fourier, Hamilton and
Maxwell, leading to successively deeper foundations of the very same
laws. In particular, the Lagrangian and Hamiltonian formulations,
which offered fresh insights into the hidden richness of the
original Newtonian premises,  paved the way to still greater depths
of knowledge  which could not possibly have been anticipated by its
Founder. Thus the Lagrangian formulation  gave birth to the concept
of  Action (as the time-integral of the Lagrangian), a new kind of
invariant which for the first time put all the four degrees of
freedom--three space dimensions and one time dimension--under one
roof. [This last concept was also to play a key role later in
the  formulation of  the Theory of Relavity --both Special and
General--at the hands of Albert Einstein, who was able to integrate
the two independent dimensions  of space and time  into an
organic whole  with the help of  a universal constant known as the
velocity of light].

\subsection {Action: A new territory}

 To come back to the virtues of `Action',
this quantity, which possesses the dimensions of angular momentum
(another key concept which was to prove vital for the feature of
\textit{discreteness} in quantum theory--see below)  in turn gave rise  to a
more  universal yet highly compact  law, called the  Principle  of
least Action, from which would naturally-and more compactly--
emerge not only the  laws of motion  from a \textit{variational principle},
but also that the latter would show  far greater predictive powers
than those realizable from the original Newtonian premises. For
example,  the Hamiltonian  equations motion-- a byproduct of the
same Principle-- which,  though identical in physical content with
the original Newtonian form, nevertheless was to show the directions
towards new territories which had hitherto remained inaccessible
to the Newtonian world.
\par
As to the "`new territory"' that had remained invisible to the
original Newtonian world, it needed the genius of Paul Dirac
[5]--inspired by Werner Heisenberg's intuitive idea of a matrix
structure [6] for the concerned dynamical variables --to replace the
classical Poisson brackets for any two dynamical variables  by the
corresponding \textit{operator commutator brackets}, obtained simply by
dividing with the Planck's constant called $\hbar$ as the basic unit
of angular momentum, together with the mysterious factor $"i"$ !
Perhaps a word about the mysterious factor $"i"$ is in order at this
stage. While its numerical value is merely a " square root of minus
one", this `static' quantity, got transformed at the hands of Dirac
to the status of a  \textit{dynamical variable} with great potential
for fresh adventures. Indeed Dirac demonstrated that this strange
quantity called "commutator bracket divided by  $i\hbar$ " happened
to  possess identical algebraic properties to the classical Poisson
brackets[5]! And this `fresh adventure' carried precisely the seeds  of
discreteness that characterizes Quantum Theory of today.
Thus was born the quantum theory, a new paradigm
emerging from the original premises of Newton's continuum theory
that would have been impossible  to guess from the Newtonian
equations of motion. Indeed, Herbert Goldstein, in his famous book on
classical mechanics [7], termed Hamilton's canonical equations as
providing  "the golden road to quantization ".  An alternative,
albeit equally revolutionary, formulation of the same paradigm of
discreteness by Erwin Schroedinger [8]-- using Louis de Broglie's
concept of wave-particle duality [9]-- gave rise to still another,
equally vibrant, form of dynamics in the shape of a wave equation
with identical physical content to the Heisenberg-Dirac form -- the
celebrated Schroedinger equation. And it took a new mathematical
vehicle --the  theory of  unitary transformations--to
prove the equivalence of the two.

\subsection{Dirac equation: A synthesis of matter and radiation}

Further incursions into  the rapidly developing territory of physics
with  the  help of   mathematical  machinery  became possible via
the realm of interaction of  (Newtonian) Matter  with (Maxwellian)
Radiation. Thus the  special  force $F$ which  characterizes
Newton's Law  for the motion of a charged particle  acquires the
form  of the \textit{Lorentz force} which  expresses the resultant of
the electric and magnetic forces on the charged particle concerned.
Conversely, the laws that determine the influence of Matter on the
evolution of the electromagnetic field could not be left far behind.
The latter have been termed Maxwell's equations, after the Man who
first gave a unified  description of the piecemeal influence of
matter on radiation,  discovered individually by several giants
(Coulomb, Gauss, Faraday, Biot-Savert, Lenz),  into one organic
whole, leading to the emergence of  \textit{light} as a (universal)
form of wave motion with an electromagnetic  origin. This mutual relationship
between the two basic entities of Nature also follows from the
Master Action Principle  defined above, as a single source of
their mutual relationship. The remarkable thing about  the
Lorentz-cum-Maxwell equations is that they are already compatible
with Einstein's  Special Theory of Relativity \textit{as they stand},
without the need for further  physical assumptions. To see Dirac's
unifying role in bringing about the synthesis of relativity with
quantum theory, his first step was to provide the quantum version
of the classical Hamilton equations through the formal
equivalence of the Poisson and Commutator brackets ( see above).
His second step [10] provided the next crucial element : a self-consistent
mathematical equation--the \textit{Dirac Equation} -- for the interaction
of a relativistic electron with the electromagnetic field, bringing out
in the process its \textit{spin} property together with  the
associated magnetic moment.

\section{Emergence of Quantum Field Theory (QFT)}

\subsection{Problems with relativity \textit{plus} quantum theory}

The Dirac equation has had a profound impact on the very direction
of physics through  its diverse ramifications born out of certain
consistency problems inherent in its formulation. The most important
one concerns the impossibility of  a self-consistent
quantum description of a single relativistic particle.  This may
sound paradoxical, since a `non-relativistic' (slow-moving) small
particle is perfectly capable of quantization, just as Einstein had
encountered no difficulty in obtaining  his`classical' relativistic
equations for a macroscopic fast-moving particle. It is only  when
both conditions (relativity and quantum theory) are  simultaneously
imposed that strange consistency problems arise. Now Dirac was
already aware of the problem of non-positive probability  inherent
in a naïve application of the second-order Klein-Gordon equation
which has two time-derivatives to go two space-derivatives so as to
preserve the structure of Special Relativity.  So he tried his luck
with first-order differential equations, viz., single time
derivative which must go only with single space-derivatives to
maintain a relativistic  balance. Such a structure had necessarily
to be at the cost of a multi-component (no longer single-component
!) wave function. [ To give a useful analogy, the Maxwell
equations  are also coupled first order differential equations in
the electromagnetic field components $E$ and $H$].  In so doing
Dirac succeeded in obtaining non-negative probability densities to
be sure, but he could not avoid the problem of negative energy
states ! He could resolve this vexing problem  only after
postulating that the vacuum (the state of lowest energy) is already
full of  negative energy states , so  that  a positive energy
electron cannot directly make a transition to a negative energy
state--thanks to the Pauli Exclusive Principle characteristic of
spin-one-half  particles. It is only when a `hole'  is created in
this sea of negative energy states by one of these negative energy
particles acquiring enough positive energy (given from outside)  to
jump out of the vacuum, that a transition to a negative energy state
by another positive energy particle  becomes possible.

\subsection{Cost of unification : concept of field}

 In making such a  hypothesis therefore, Dirac  was  dealing effectively with,
not one but, an infinite number of particles (a field) at the same
time !   And he was promptly vindicated by Anderson's cosmic ray
discovery of a positively charged particle with the electron's mass,
playing precisely the part of this `hole' !  Pauli  immediately saw
through the  `message'  of  Dirac's negative energy sea
interpretation , and proceeded to show that  this property of
spin-half electrons (known as Fermions) being accompanied by an
infinite number of its kind  is also shared by the spin-zero
particles (known as Bosons) which obey the  second order
Klein-Gordon equation. All that was needed was that  the parameter
of \textit{energy}  be replaced by the parameter of \textit{probability} density
[11]. Thus the non-positive probability implicit in a Klein-Gordon
wave function should now be re-interpreted as an average charge
density of  an infinite number of spin-zero particles--a field
again--of both positive and negative charges ! Thus the common
message from both cases is that of the existence of an \textit{infinite}
number of particles --be it spin-half Fermions or spin-zero Bosons--
so as to be consistent with  relativistic quantum theory. A single
particle just will not do when quantum theory and relativity.are
sought to be put together.  It was perhaps Freeman Dyson who,
through  his Cornell lectures of 1951 [12], clearly brought out the
field role for both fermions and bosons, in preference to  their
single particle interpretation.

\subsection{Mathematical language of QFT}

 This in short is the story of the  genesis of Quantum Field Theory,
 or QFT for short,
 whose applications in physics extend all the way  from the theory of
 elementary particles  to condensed matter physics, and now spreading
 even to the biological sciences ! Its  mathematical language being
 that of the \textit{harmonic oscillator} (HO) ,  QFT may be regarded as a
 collection of harmonic oscillators  which  must be systematically
 classified and put to different jobs. Now for a single HO,
 the total energy is half kinetic and  half potential, so  it is
 convenient to define its basic variables  as  50 -50 ad
 mixtures of  momentum $(p)$ and coordinate $(x)$  variables
 whose basic commutator $[x, p]$  has the value $i\hbar$ by virtue of their
 Poisson bracket structure (see Dirac above). These 50-50 admixtures,
 termed  $a$ and $a^\dag$,  expressed in dimensionless units,  may be
 easily shown to obey the  commutation relations
 $$ [a, a^\dag] =1; \quad [a, a] = [a^\dag, a^\dag] = 0.  $$
 And  the energy operator $H$ which equals  $\omega \hbar (a^\dag a +1/2 )$,
 in units of  the spring constant $\omega$,  incorporates the essential
 dynamics in terms of the `number operator' $N$ which equals $a^\dag a$,
 and whose  integer eigenvalues
 $n$ are called  occupation numbers; the corresponding `states' (wave functions)
 are called the eigenstates $|n>$ of  $N$.  By  virtue of the  equalities
 $a | n >$= $\sqrt{n} | n-1 >$ and   $a^\dag | n >$ = $\sqrt{n+1} | n+1 >$,
 $a$ and $a^\dag$ are called destruction / creation operators  respectively,
 since they reduce / increase the occupation number in a given state $|n>$ by
 one unit each.  Hence by successive applications of the $a^\dag$ operators, the eigenstates
 $|n>$  can be built up from  the  ground state (`vacuum')  $| 0 >$.  In this language
 the successive eigenvalues of the energy operator become  $E_n$ which equals $\omega\hbar ( n + 1/2)$ .
 The QFT generalization for an infinite collection of  harmonic oscillators
 $a_k$, $a_k^\dag $ indexed by an integer $ k $(or a collection of such integers thereof) ,
 is now  only a matter of  systematic  construction  of the energy contents
 as well as of the  successive states which are all expressible in terms of a "master" ground state $|0>$.
 This " master"  ground state -- or  simply the  \textit{Vacuum state} -is a central  theme
 around which  the entire  concept and  methodology  of QFT  devolve.

 \subsection{Applications of QFT: success of QED}

The earliest application of  the QFT formalism has been  in the area
of quantum electrodynamics -QED for short-the theory of interaction
of charged fermions (electrons, free or bound) with the bosonic
electromagnetic  field,  which lends itself easily to an HO
formulation. And the  success of QED  in unravelling the mysteries
of  the `fine-structure constant'  for understanding atomic and
molecular spectroscopy (upwards of the Lamb Shift)to the accuracy of
"`one in a trillion"' is but too well known for further elaboration [13]. Even Dirac  was
impressed by this achievement, in answer to Dyson's query,  but he
did not feel happy at this development,  and wished the formalism
were not so "ugly" [14] !

\section{Conclusion : feedback effects}

This long story of the strong interaction of mathematics with
physics, which suggests a sort  of mutual interdependence, need not
give the impression that  the influence of Mathematics on the other
physical sciences is any less profound,
so as not to invalidate in any manner the basic theme of this essay.
a whole. Indeed  mathematical techniques are  finding increasing
applications into most other  sciences (from the physical to the
biological); only the aspect of  mutual interdependence that has
characterized its relationship with theoretical physics, is  perhaps
absent !

Apart from  the influence that Mathematics exerts on the Sciences,
its own frontiers are expanding  daily in newer and newer
directions. Perhaps the most significant  development in this regard
is the expansion of the techniques of Mathematics to the domain of
\textit{information theory} as well as  the  related field  of \textit{computer
technology}. In this respect, a vital ingredient of physics, namely
quantum theory  has had  a crucial role,  inasmuch  as  `quantum
computation'  -- still under active development--has a great potential
for  a significantly faster action  than  its  classical counterpart.

Before ending  this narrative  on the role of Mathematics in shaping
the different Sciences, it is perhaps in order to express some
thoughts on the nature of the \textit{feedback} from the latter to the
former. Namely, how does Science as a whole react to the language of
Mathematics ? To address this question,  one may wish to inquire
into the methodology of Science in actual practice, namely the
scientific method.  In this respect the vital role of checking or
verification is crucial . To support this view,   it is of interest
to quote from  P. W. Bridgeman ,  taken from  his Nobel Lecture
(1955) [15]:
 " Scientific method is something talked about by people
standing on the outside and wondering how the scientist manages to
do it. These people have been able to uncover various generalities
applicable to at least most of what the scientist does, but it seems
to me that these generalities are not very profound, and could have
been anticipated by anyone who know enough about scientists to know
what is their primary objective. I think that the objectives of all
scientists have this in common--that they are all trying to get the
correct answer to the particular problem in hand. This may be
expressed in more pretentious language as the pursuit of truth. Now
if the answer to the problem is correct there must be some way of
knowing and proving that it is correct--the very meaning of truth
implies the possibility of checking or verification. Hence the
necessity for checking his results always inheres in what the
scientist does."

This essay is dedicated to the memory of my Father, Jatindranath
Mitra, who had been the main inspiration  in  my pursuit of
Mathematics.

\end{document}